# Barkhausen Noise Probe of the Ferroelectric Transition in the Relaxor $PbMg_{1/3}Nb_{2/3}O_3$-12% $PbTiO_3$

Xinyang Zhang[1], Corbyn Mellinger[1,2], Eugene V. Colla[1], and M. B. Weissman, and D. D. Viehland[3]

[1] Department of Physics
University of Illinois at Urbana-Champaign
1110 West Green Street, Urbana, IL 61801-3080

[2]Department of Physics and Astronomy,
University of Nebraska-Lincoln,
Lincoln, NE 68588.

[3]Department of Materials Science and Engineering,
Virginia Tech,
Blacksburg, Virginia 24061

**Abstract:**

Barkhausen current noise is used to probe the slow field-driven conversion of the glassy relaxor ferroelectric state to an ordered ferroelectric (FE) state. The frequent presence of distinct micron-scale Barkhausen events well before the polarization current starts to speed up shows that the process is not a conventional nucleation-limited one. The prevalence of reverse switching events near the onset of the rapid part of the transition strongly indicates that electric dipole interactions play a key role. The combination of Barkhausen noise changes and changes in the complex dielectric response indicate that the process consists of an initial mixed-alignment domain formation stage followed by growth of the domains aligned with the applied field.

## Introduction

Relaxor ferroelectrics have a large metastable regime, in which a disordered glassy paraelectric state (RXF) persists for unmeasurably long times so long as the applied electric field E is small [1, 2], although the thermodynamically favored state is ferroelectric (FE) even at E=0. [2, 3] This metastable regime is illustrated in Fig. 1 for a typical cubic perovskite relaxor, $(PbMg_{1/3}Nb_{2/3}O_3)_{0.88}(PbTiO_3)_{0.12}$, which we call $PMNPT_{12,}$. [4] Since many practical applications of relaxors rely on the high ac piezoelectric coefficients that are lost on conversion to an ordered FE state [4] there are practical reasons to explore what governs the kinetics of the conversion from RXF to FE. This process is also of interest as one of the more experimentally accessible examples of conversion from glassy to ordered states [5], thanks in large part to the current $I_P(t)$ that accompanies that transition when the sample is configured as a capacitor.

The conversion to FE after application of E involves a period in which the polarization creeps up, followed by a fairly abrupt transition to FE, typically showing one or two peaks in $I_P(t)$. Previous studies of the E and T dependences of the RXF to FE conversion kinetics in $PMNPT_{12}$ (and other compositions) showed that just below the true thermodynamic melting temperature of the FE phase the divergence of the characteristic lag-time before the rapid parts of the transition did not fit the standard picture of nucleation of an ordered phase from a homogeneous glass. [6] The form of the discrepancy suggested that the geometry of nucleation sites might not be compact balls but perhaps something more like fractal clusters. [6] Nevertheless, curve-fitting of conversion times vs. (E, T) provides little guidance in picturing the kinetic pathway.

Much more detailed information can be found in the Barkhausen noise in the conversion process. Barkhausen noise in $I_P(t)$ can be measured at fixed E, experimentally much simpler than disentangling Barkhausen noise from deterministic polarization response taken while E is changing.[7] In this paper we report that Barkhausen noise in $PMNPT_{12}$ reveals several key features of the FE formation process. The noise shows that micron-size FE nuclei often form in the creep stage well before the main conversion event, showing that the basic assumption of a nucleation-limited rate is wrong. Just before the start of the rapid part of the transition, there are typically some reverse-oriented polarization spikes, which we shall argue imply that long-range electric dipole interactions play a key role in limiting the conversion rate. That conclusion will have major implications for the geometry and growth kinetics of the initial FE regions. Combining the Barkhausen patterns with information from the time-dependent dielectric response function will provide strong indication of the distinct processes [8] that occur in the initial and final stages of the conversion to the FE phase. In addition, we shall note a variety of other noise effects that shed less direct light on the underlying physics and raise questions for future study.

**Materials and Methods**

The $PMNPT_{12}$ samples studied here are from the same batch previously described in our work on the FE formation kinetics. [6] The samples of PMN-12%PT, provided by TRS Technologies (State College, PA), were grown by a modified Bridgman technique. The samples were configured as parallel-plate capacitors oriented with the applied **E** along a [111] axis, an easy polarization axis for the FE state. The PMN-PT sample was initially ~0.48mm thick with ~1mm$^2$ electrode area on a larger sample. It was slightly thinner after repolishing for some later data. Contacts were made via evaporated Ag layers of roughly 200nm thickness on top of adhesion-enhancing ~10 nm thick evaporated Cr layers. The measurement circuitry used externally fixed voltages on the sample and a current-sensing amplifier.

The transition to the FE phase was monitored using the pyroelectric current $I_P(t)$ and the complex dielectric response function, ε'-iε". In order to reduce electromagnetic pickup the nitrogen-flow transfer-line cryostat was mounted inside a double-wall mu-metal shield. The shield was supported on a sand pile to reduce vibrational pickup. Measurement of $I_P(t)$ used a fixed capacitance voltage with a first-stage very-low-noise op-amp current amplifier circuit, shown in Fig. 2. The op-amp and the following instrumentation amp, as well as their lead-acid battery power supply, were mounted within the mu-metal shield. The output signal was then amplified on two separate channels. One, first sent through a low-pass filter usually set at 20Hz, is measured by a DMM at a 10Hz sampling rate, giving our low-frequency (LF) signal. The LF signal is what would usually be monitored to follow the transition. The other is sent through an ac-coupled high-pass filter usually set at 1 Hz, and then a low-pass anti-alias filter set at 2.1k Hz before sampling at 5kHz by a 16-bit A/D, giving our HF noise signal. HF noise runs were taken during periods when the lab was unoccupied by other users, to avoid electrical and mechanical pickup from other experiments. In some cases, rather than make the HF measurement we applied an ac voltage of 28.3 mV rms at 100 Hz, to measure ε' and ε".

Since, as discussed below, a distinct non-Barkhausen low-frequency noise source obscured data below about 10 Hz, we usually used a detrending algorithm on the HF data to avoid artifacts of the LF signal and noise appearing in the HF data, especially due to the form of the discrete Fourier transform used to take spectra. The detrending procedure removed a least-squares third-order polynomial fit from 500-point (0.1s) windows. 60 Hz and harmonics were removed by digital filtering, using an FFT and inverse FFT.

We followed the RXF→FE conversion under several protocols. Most commonly, a "ZFC" procedure was used, in which the sample was annealed at 450K, then cooled at a rate of 4K/min with E=0 to a test temperature in the hysteretic range, at which point a field was applied. T and E were adjusted so that the conversion time was typically in the range of several hundred seconds. In other cases ("FC"), we followed the process during cooling at 4K/min in a fixed field.

**Results**

Fig. 3 shows a typical example of the LF $I_P(t)$ along with the changes in ε' and ε" after application of a field after a ZFC protocol. The pattern is similar to those found previously. [8] An initial $I_P(t)$ pulse occurs when E is applied, gradually decaying to a lower creeping $I_P(t)$. After a delay time, $I_P(t)$ rises again as the sample goes FE. In the sample used here in under the conditions most convenient for the Barkhausen measurements, the rapid transition appears as an asymmetrical broad peak in $I_P(t)$ followed by another very slow creep as the sample approaches saturation. In some temperature and field regimes, this peak turns into two clearly resolved peaks, with the first peak in $I_P(t)$ accompanied by a spike in ε". For the single peak here, the spike in occurs in the earlier part. Subsequently, both ε' and ε" decrease substantially as the FE phase forms. In some regimes the two peaks overlap too much to clearly distinguish them.

Subsequently setting E=0 gives a small negative pulse in $I_P(t)$. On warming the sample then depolarizes with a clear transition to the RXF (sometimes in a single step, sometimes in a bimodal step), followed by a small further loss of polarization on further heating.

One puzzling feature, observed before [3], is a large LF noise whose magnitude approximately tracks the net polarization rather than the average current or applied voltage. Microphonics would be an obvious suspect for such a polarization-dependent noise in samples that become piezoelectric when polarized. Turning off the nearby vacuum pump, the main source of vibrations, had no noticeable effect on the LF noise. Deliberate tapping on the transfer line or mu-metal shield had little effect. The LF noise showed peculiar T-dependence, getting very large at very low T and consistently showing a minimum amplitude at about 282K in our original sample. Most importantly, it also appeared under applied E at temperatures not only above the FE regime but even above the peak in ε'(T), thus ruling out any direct simple relation to the FE state or even to the slow kinetics of the RXF state. We suspect that this LF noise may involve microphonic response to large-scale strain relaxation, but shall not discuss it further until more complete evidence is obtained. The LF noise is irrelevant to the HF Barkhausen noise features to be discussed here, except that it unfortunately limits the study of the Barkhausen spectrum during the initial creep phase.

The key features of the HF noise on a typical ZFC protocol are illustrated in Fig. 4, which shows LF $I_P(t)$ along with HF $I_P(t)$. Fig. 5 shows expanded views of individual spikes at various stages of the process. Several of these features are worth noting.

1) There are unmistakable Barkhausen spikes early in the creep stage. This result is not universal for all the data taken, but it is typical. (Such spikes also appear in field-cooled runs well before the main transition.) The integral of the current under a typical distinct spike (e.g. 5a) in this stage gives a net charge corresponding to polarization of about 30 $\mu m^3$ of the sample, more than sufficient to constitute a nucleation site in any standard nucleation picture.
2) Closer inspection of the regime just before the main rapid FE transition shows several Barkhausen spikes in the reverse direction of the net polarization, as illustrated in the expanded view in Fig. 6. This figure shows that these reverse spikes are not a rare fluke but in fact are usually observed. Their significance will be discussed later. Even when, as in some data on a repolished, re-annealed, and re-contacted sample, the distinct backward spikes are not easily identifiable, their presence can be inferred from a change in skewness of the Barkhausen noise.

3) There is an abrupt onset of continuous Barkhausen noise during the rapid part of the polarization transition.
4) Clear Barkhausen spikes persist after the main rapid transition, in the regime in which the polarization slowly approaches saturation.
5) Some clear depolarization spikes occur not only during the main melting transition but also after it as the polarization approaches zero.

The onset of continuous Barkhausen noise suggests a regime in which aligned domains are growing at the expense of anti-aligned domains, similar to the conditions under which ordinary Barkhausen noise is observed. [9] One would expect that the spectrum in that regime would be approximately of the form typical for most Barkhausen noise [9, 10], rolling from flat or increasing below some characteristic frequency to roughly $1/f^2$ at higher frequency. Fig. 7 shows a spectrum taken from this regime, with and without background noise subtracted. The background is taken from the same records after the continuous Barkhausen noise has died down. This subtraction is only approximate since some Barkhausen noise remains in the background, the dielectric coefficient has changed slightly at that point, and the low-frequency noise has increased slightly due to increased polarization. Still, in the middle of the frequency range, where these background effects are least important, the spectrum does have the expected form. In the range from 16 Hz to 256 Hz the background subtraction has little effect, as does the detrending procedure, so the flattening of the Barkhausen spectrum below ~100 Hz is real. The precise form of the roughly $1/f^2$ regime could be affected by the Barkhausen component in the subtracted background.

During the initial rapid stage of the transition, before the onset of the continuous Barkhausen noise, initial inspection shows no obvious Barkhausen events other than the negative spikes.

Since the onset of continuous Barkhausen noise is likely to be connected with domain wall motion, it is important to check how closely it corresponds to the time at which the spike in ε" occurs, since ε" also should be maximal around the time when the domain wall area is maximal. Although it is difficult to take HF noise data and ε" data in the same run, the corresponding times can be matched up well because each type of data is taken along with LF $I_P(t)$, whose characteristic structure serves to keep track of the timing of the FE formation process. Fig. 8 shows HF noise and ε" (t), aligned via the $I_P(t)$ data. The negative Barkhausen spikes appear during the period when ε"(t) is gradually increasing. The onset of continuous Barkhausen noise occurs later as ε"(t) is still increasing. The peak in ε"(t) roughly coincides with the most intense continuous Barkhausen noise. (There are also some poorly reproducible periods of intense Barkhausen noise later in the process, not corresponding to any clear features in ε"(t) or $I_P(t)$.) Qualitatively similar results were found also in a similar experiment at 250K.

**Discussion**

The Barkhausen noise provides a plethora of data on the pathway from RXF to FE. Some of the implications of that data are already clear and important. Others require some speculation at this point but allow us to fill in more details of a tentative picture. The two most important results here are that micron-scale Barkhausen events often occur well before the main cooperative

RXF→ FE transition and that near the start of that transition there are typically some Barkhausen events with polarization opposite to that favored by the applied field.

The presence of micron-scale Barkhausen events well before the rapid part of the transition implies that the delay between the application of a field and that rapid transition cannot be modeled as a nucleation-limited process, despite the rough qualitative agreement of the temperature and field dependences of the delay time with simple homogeneous nucleation theory. [5] Often, there are already nucleation sites present even as the polarization creep is slowing down after the initial field switch. The previous conclusion[6] that the divergence of the delay time near the thermodynamic transition temperature required some qualitative modification of the homogeneous nucleation picture thus turns out to have been understated.

The backwards spikes near the start of the main transition provide a valuable clue as to what limits the rates. These spikes must be driven by an electric field pointed opposite to the applied field and extending over at least micron-scale dimensions. The obvious origin for such a field is the dipole field lateral to FE regions that are properly aligned with the applied field. Thus these backwards spikes are a forceful reminder of the importance of the electric dipole interaction, one of the ingredients not included in simple nucleation pictures nor in our previous speculations on the limits of that picture.

The backwards spikes are inconsistent with one previously plausible picture of the creep stage. If the creep consisted of gradual alignments of uniformly dispersed polar nanodomains with the applied field, the internal field would be uniform on a large scale. Thus there would be no micron-scale regions coherently aligning in the opposite direction. Therefore that sort of mean-field picture of the creep phase is ruled out.

We may speculate a bit on how the electric dipole interactions affect the kinetics. We have tuned E and T here so that the polarization grows perceptibly but slowly. The growth rate is very field-sensitive. Say that a small region has converted to properly-aligned FE. At its ends along the field direction, the field in the neighboring RXF will be strongly enhanced compared to the applied field. The field laterally outside the FE region will be pointed the opposite direction, and not quite as strong. Growth should then be fastest along the field direction, leading to fibers of properly-aligned FE phase. These 1-D fibers should then grow at approximately constant rates since the growth area at the ends is independent of length, unlike 3-D growth with an expanding 2-D growth surface.

Thus we speculate that the creep phase consists primarily of this growth of FE fibers. This sort of 1-D growth pattern would also go a long way toward explaining the very large metastable RXF regime. Relaxors have large random fields from the compositional disorder. [11] One dimensional FE growth will be blocked when each end runs into sufficient fields disfavoring at least that particular vector polarization alignment.

Once macroscopic FE fibers are formed, there will be large lateral regions with opposite-pointing E. Some of these will flip cooperatively to the FE state, producing the negative Barkhausen spikes. At this point, the sample is converting to predominately FE, but with mixed domain orientation. The point at which most of the sample is FE but consists of large numbers of small domains with lots of domain wall area should lead to a maximum in ε", the component of the dielectric response most sensitive to domain wall motion. Just such a spike is consistently seen at this point of the process. [8]

Once the sample is filled with FE domains, the free energy can be lowered by growth of the properly aligned domains at the expense of the anti-aligned domains. That should lead to a fairly typical Barkhausen signal, with a spectrum turning from an increasing function of f to a roughly $1/f^2$ form, as we typically saw.

The Barkhausen noise sheds light not only on the initial stages of the RXF→FE transition but also on the reason that this transition often occurs in multiple stages, most typically in two distinct stages in fresh samples with sharp transitions.[8] Our speculations on these two stages previously had considered different roles for the realignment of polar nanodomains with different starting orientations, roles for the collapse of inter-nanodomain glassy structure, or perhaps an intermediate symmetry (e.g. tetrahedral) between the cubic and rhombohedral phases.[8] The Barkhausen noise does not rule out any of those ideas but does suggest an entirely different picture. The first stage on the RXF→FE freezing path and the second stage of the FE→RXF melting path are typically sharp, with characteristic temperature for FE→RXF not strongly dependent on the sample history. The second stage of freezing, corresponding to the first stage of melting, is often broader and more sensitive to prior aging in field and other aspects of sample history.

Based on the Barkhausen noise picture described above, we think it is most plausible to attribute the sharp step to the overall conversion from cubic to rhombohedral, expected to produce a sharp cooperative transition thanks to long-range strain interactions. The noise indicates that the broader, more history-dependent step consists at least in part of conversion from a state of mixed aligned and anti-aligned domains to a state of almost completely aligned domains. That such an intermediate mixed-alignment domain state spontaneously reforms on reheating with E=0 is consistent with the role of strong local fields resulting from quenched disorder. These fields apparently make the mixed-domain state thermodynamically favored just below the rhombohedral phase transition, at least for small applied E.

In summary, the Barkhausen noise in the conversion of the glassy relaxor state to the ordered ferroelectric state reveals that the slow kinetics are not nucleation limited. The electric dipole interaction plays a key role in slowing the process. That interaction may lead to the initial formation of fibers rather than blobs of the FE state, an interesting issue for theoretical investigation. The changes in the noise properties suggest that the initial stage of the conversion involves formation of rhombohedral domains along the field axis but of mixed polarization direction, followed by a stage in which domains with the field-favored polarization direction grow.

**Acknowledgments:** Corbyn Mellinger was supported by NSF REU grant 13-59126. X. Zhang was funded by the John A. Gardner Undergraduate Research Award. We thank Owen Huff, also supported by NSF 13-59126, for technical assistance and useful conversations, and thank the UIUC Physics Department for support.

## References

[1] B. Dkhil and J. M. Kiat, J. Appl. Phys. **90**, 4676 (2001).

[2] E. V. Colla, N. Jurik, Y. Liu, M. E. X. Delgado, M. B. Weissman, D. D. Viehland, and Z.-G. Ye, J. Appl. Phys. **113**, 184104 (1 (2013).

[3] E. V. Colla, D. Vigil, J. Timmerwilke, M. B. Weissman, D. D. Viehland, and B. Dkhil, Phys. Rev. B **75**, 214201 (2007).

[4] L. E. Cross, Ferroelectrics **76**, 241 (1987).

[5] V. M. Fokin, E. D. Zanotto, N. S. Yuritsyn, and J. W. P. Schmelzer, J. Non-Crystalline Sol. **352**, 2681 (2006).

[6] E. V. Colla, J. R. Jeliazkov, M. B. Weissman, D. D. Viehland, and Z.-G. Ye, Phys. Rev. B **90** (2014).

[7] E. V. Colla, L. K. Chao, and M. B. Weissman, Phys. Rev. Lett. **88**, 017601 (2002).

[8] E. V. Colla and M. B. Weissman, Phys. Rev. B **72**, 104106 (2005).

[9] G. Durin and S. Zapperi, J. Magn. Magn. Mat. **242-245**, 1085 (2002).

[10] M. C. Kuntz and J. P. Sethna, Phys. Rev. B **62** (2000).

[11] B. P. Burton, E. Cockayne, and U. V. Waghmare, Phys. Rev. B **72**, 064113 (2005).

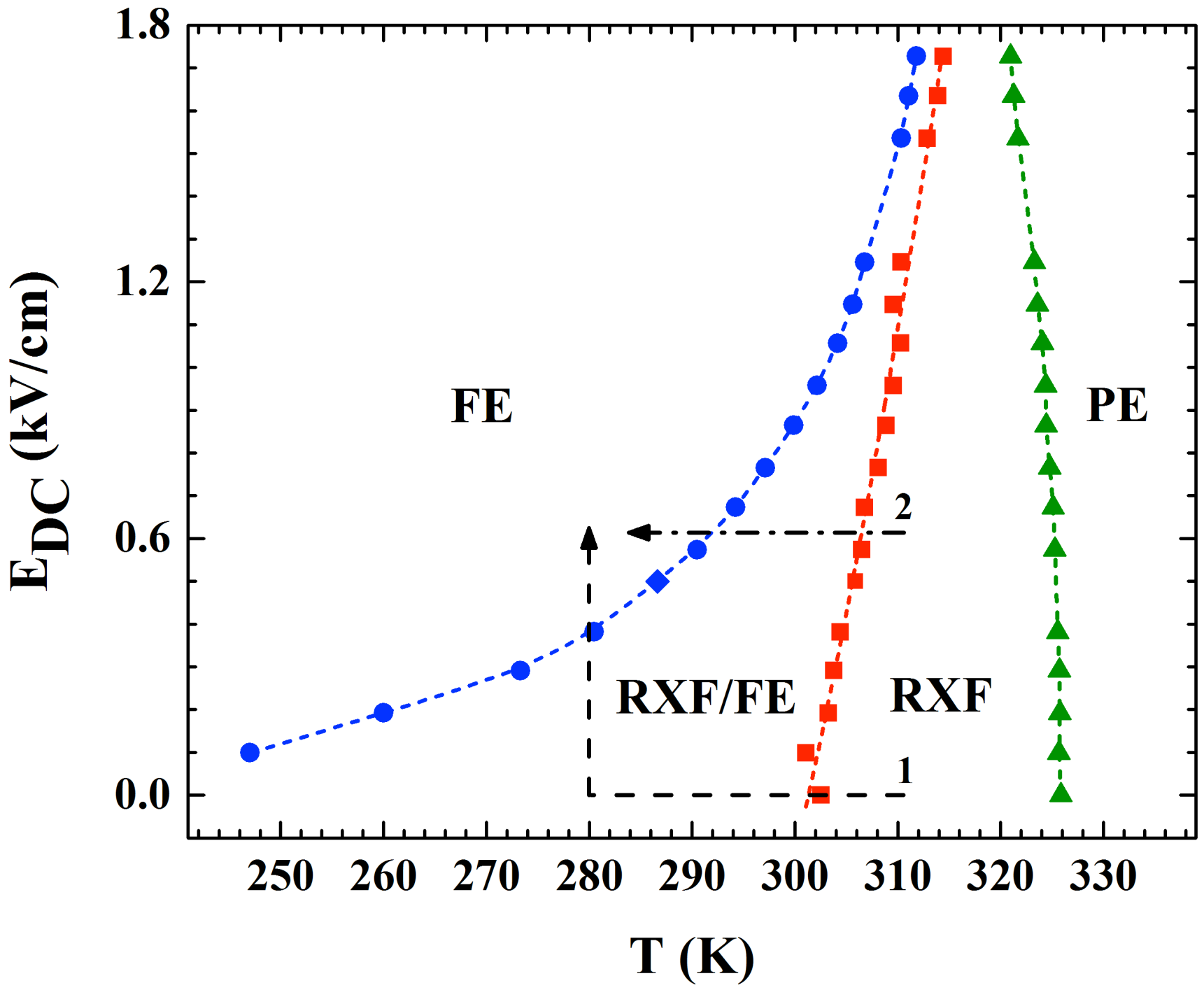

Figure 1. The empirical phase diagram of a PMNPT12$_{12}$ sample form the same batch as those used for the noise studies shows the equilibrium paraelectric phase (PE), separated by a frequency-dependent crossover from the non-equilibrium relaxor state (RXF). The RXF is separated from the FE state by a nearly rate-independent melting line and a somewhat rate-dependent freezing line. The large gap between the melting and freezing temperatures at low E gives the large hysteretic regime. Path "1" shows a ZFC process and path "2" shows an FC process.

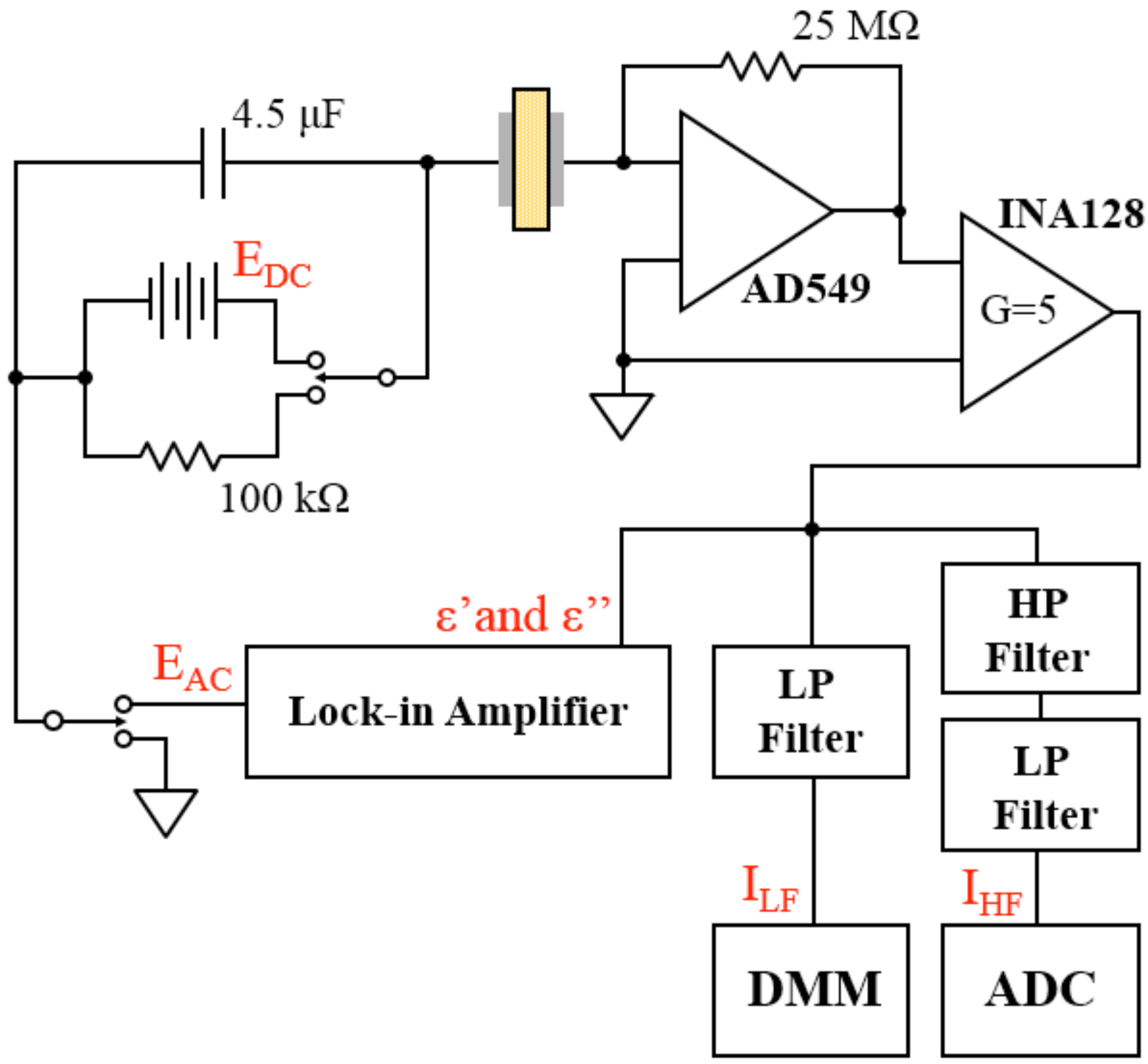

Figure 2. The block diagram of the circuit for measuring the LF and HF components of the polarization current, $I_P(t)$. The key component is the initial current-to-voltage converting op-amp AD549, with very low current noise, which is followed by instrumentation amp INA128. The sample, located at the AD549 input, is shown as a slab with shaded contact plates.

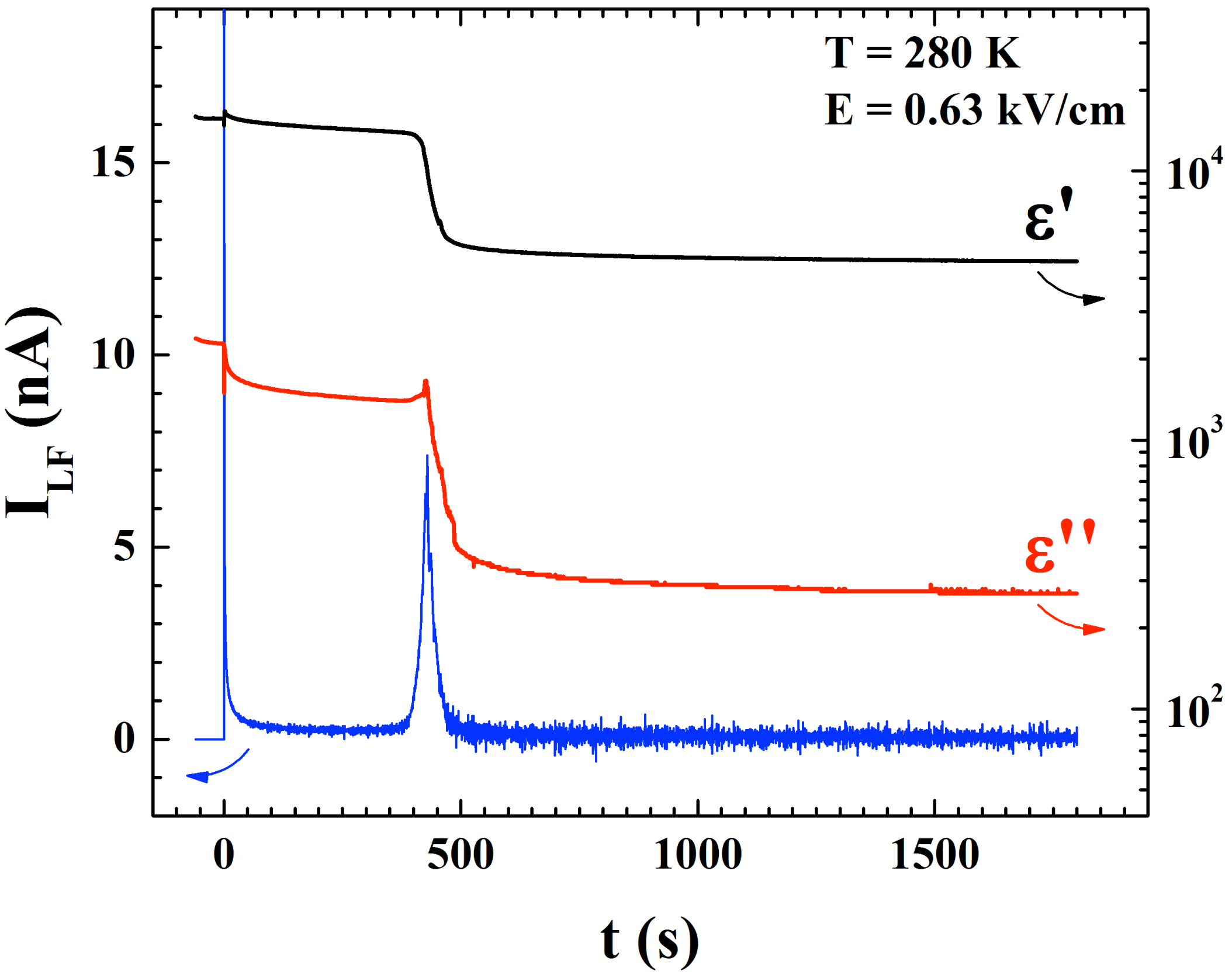

Figure 3 shows a typical example of the LF $I_P(t)$ along with the changes in ε’ and ε” on a ZFC protocol.

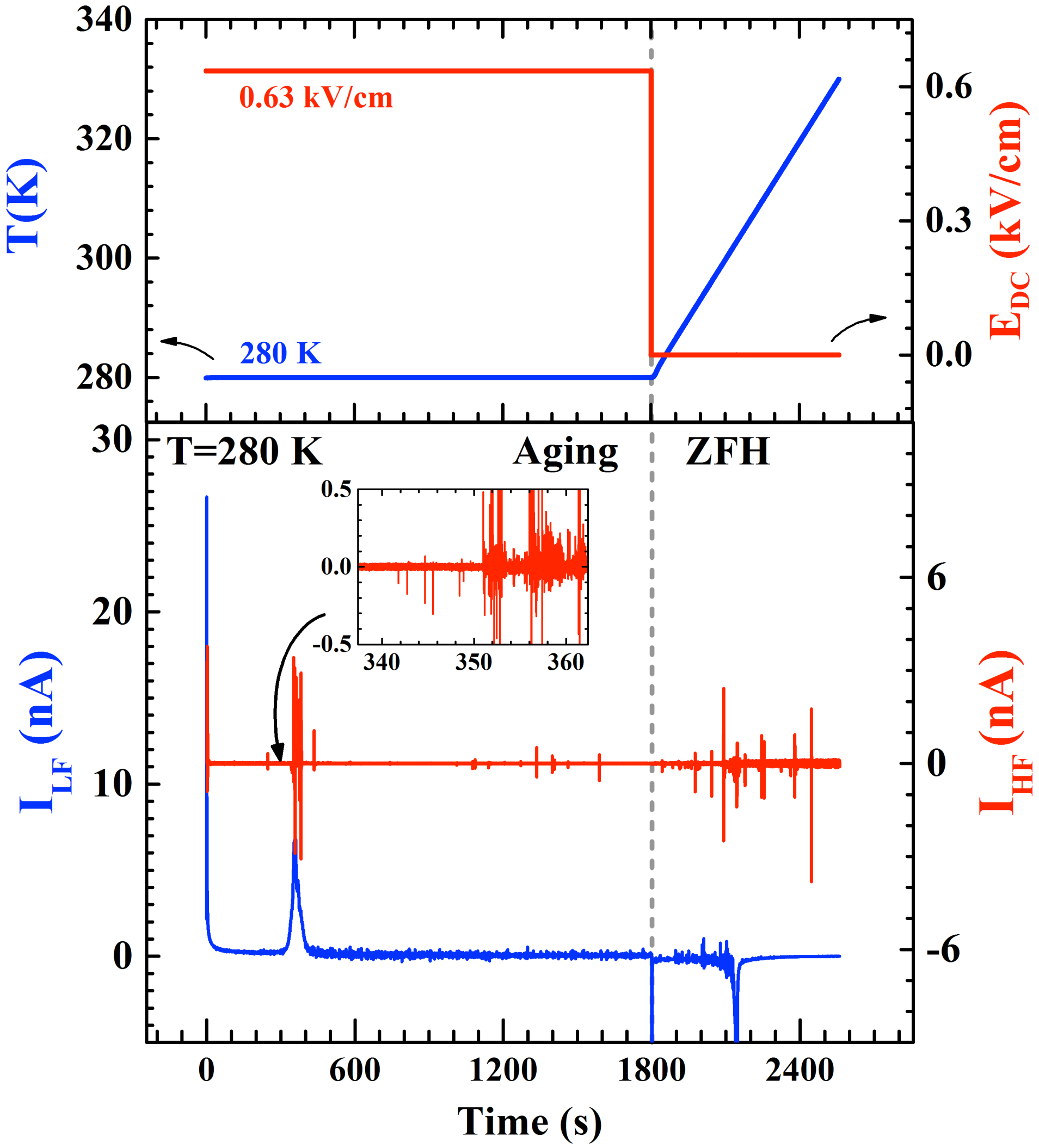

Figure 4. The key features of the HF noise on a typical ZFC protocol are illustrated in Fig. 4, which shows LF $I_P(t)$ along with HF $I_P(t)$.

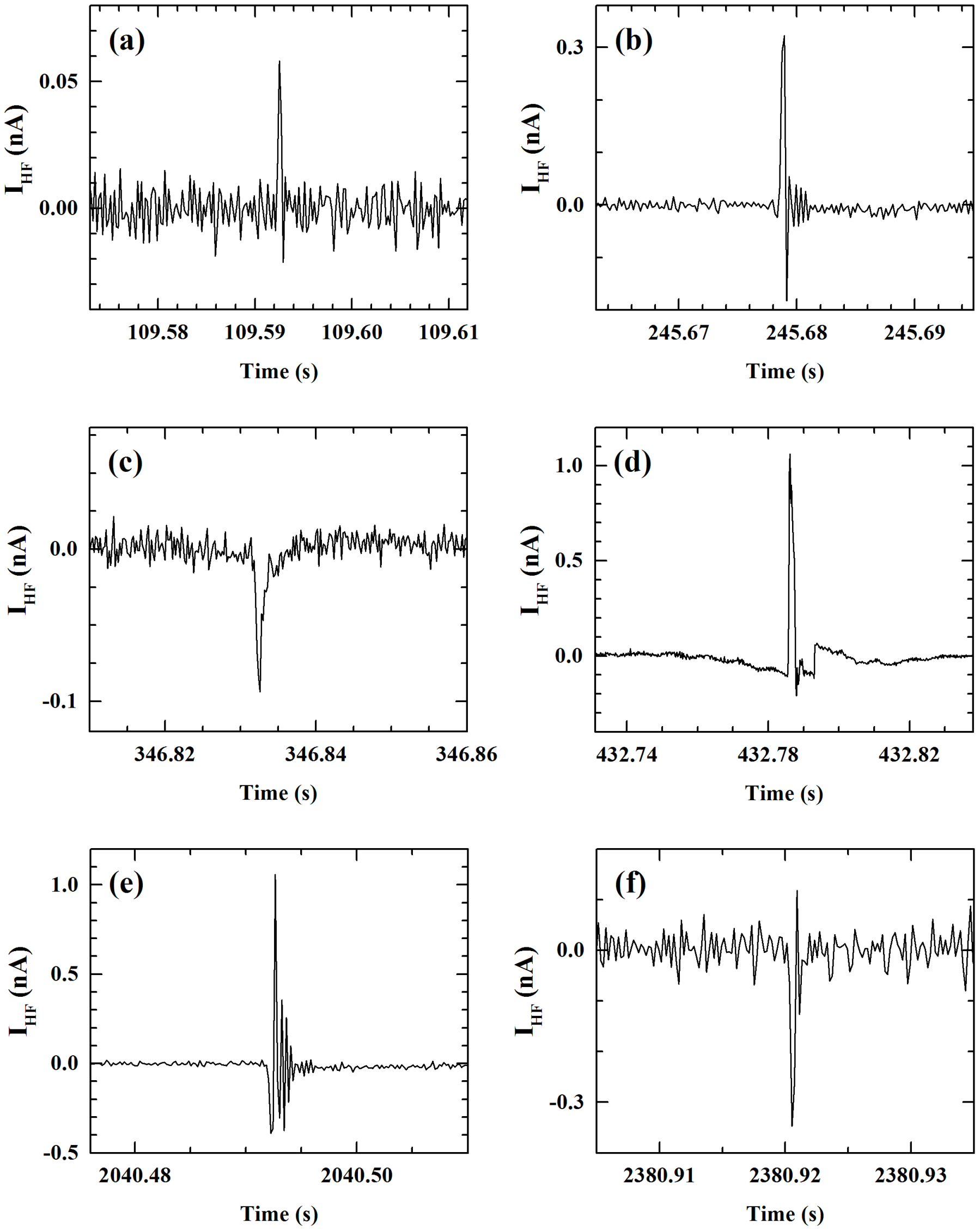

Figure 5 shows expanded views of individual spikes at various stages of the process. (a) creep stage (b) creep stage (c) late creep stage (d) after FE transition (e) after setting E=0, before melting (f) after main melting. The ringing seen most clearly in (b) is the response of the anti-alias filter to a brief pulse. Longer pulses, e.g. (c), smooth out the ringing. Pulse (e) appears to come from a mixed pair of positive and negative pulses.›

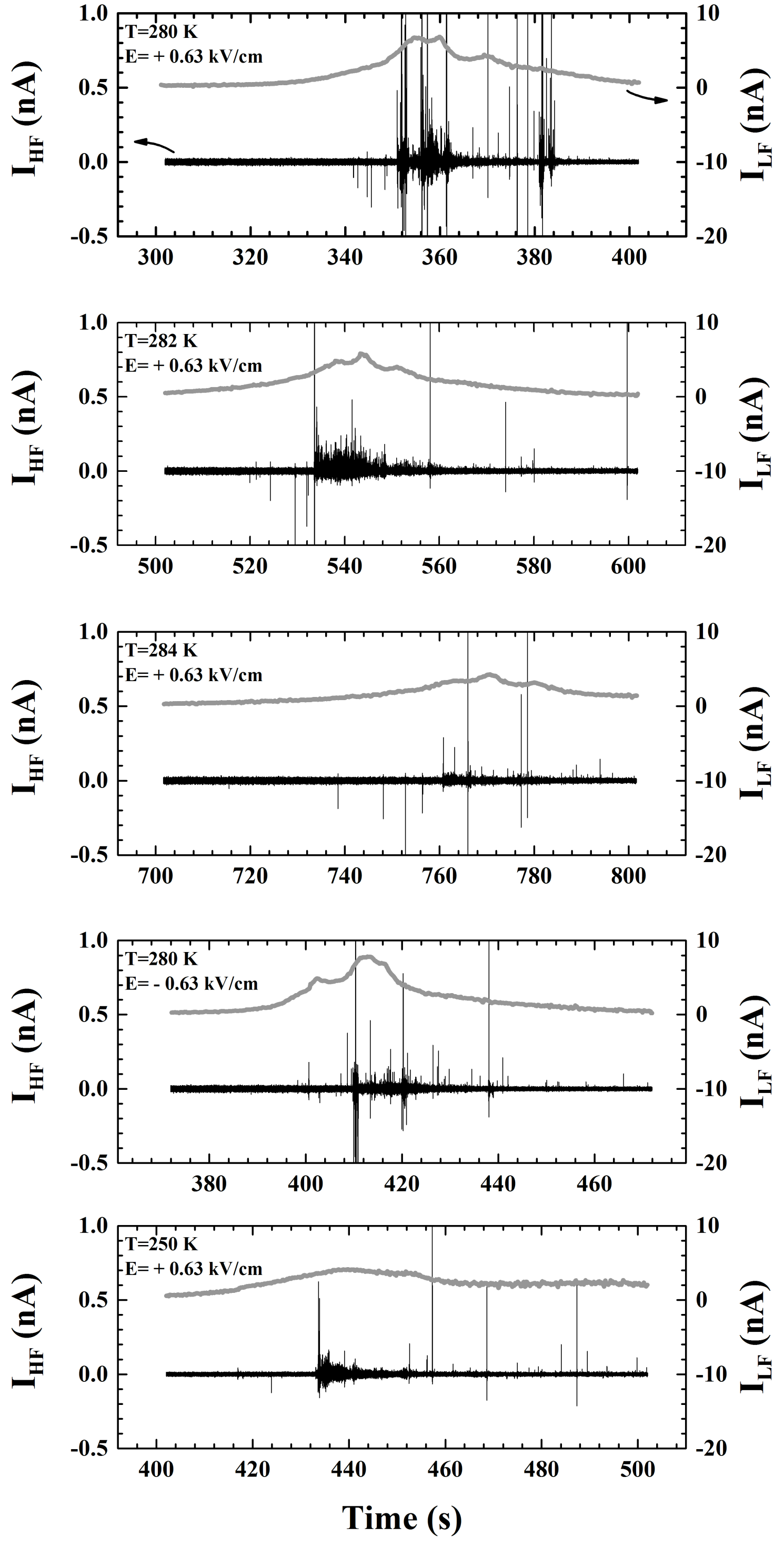

Figure 6 shows that Barkhausen spikes in the reverse direction of the net polarization, are typically found in the early stages of the main RXF→FE conversion event..

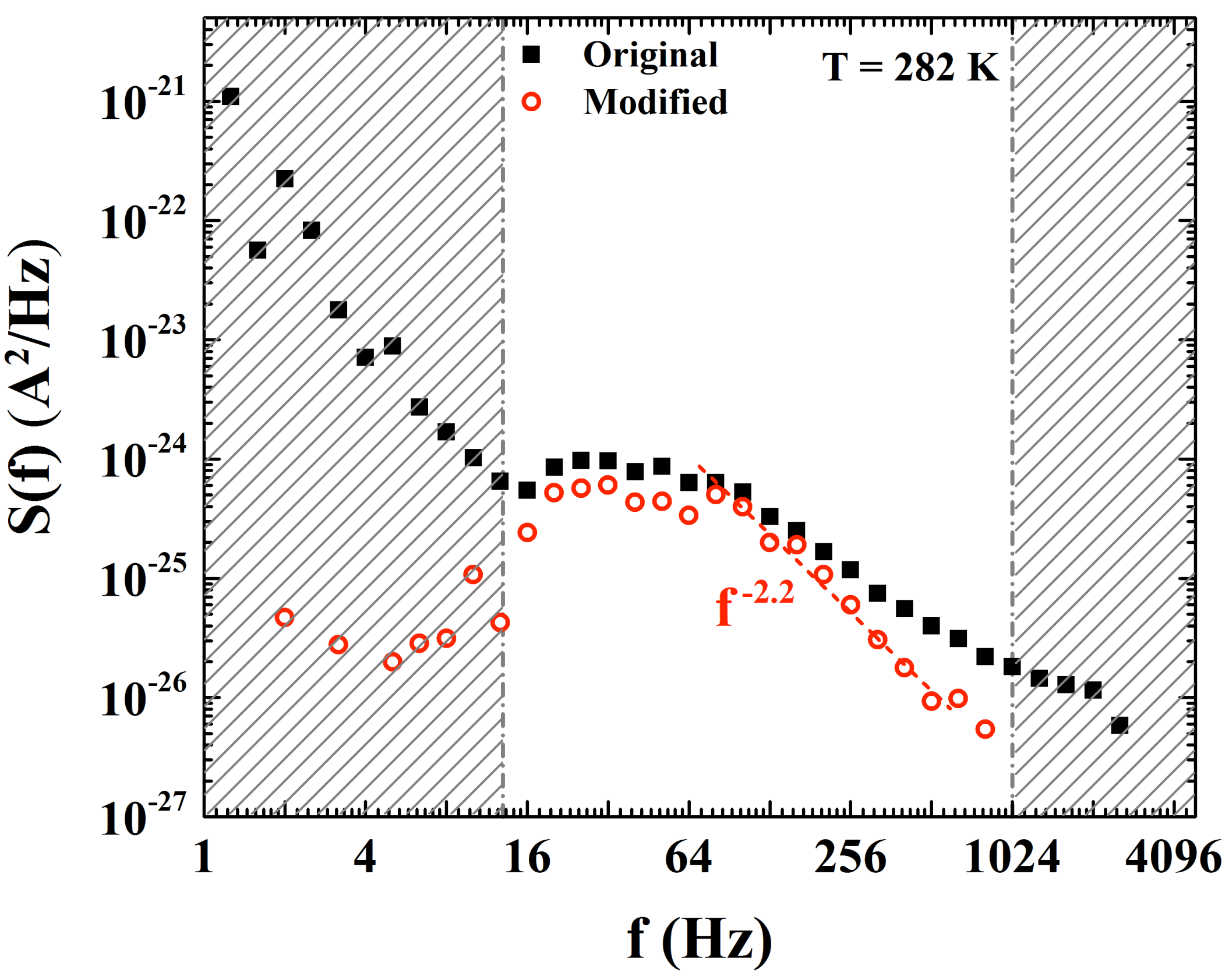

Fig. 7 shows a spectrum taken from the regime of continuous Barkhausen noise. The squares are the bare spectrum. The circles show the same data after detrending and subtracting the spectrum of a detrended background taken just after the Barkhausen noise largely subsided. The shaded areas indicate regions in which the uncertainty in the time-dependent background subtraction prevents reliable estimates of the Barkhausen spectrum.

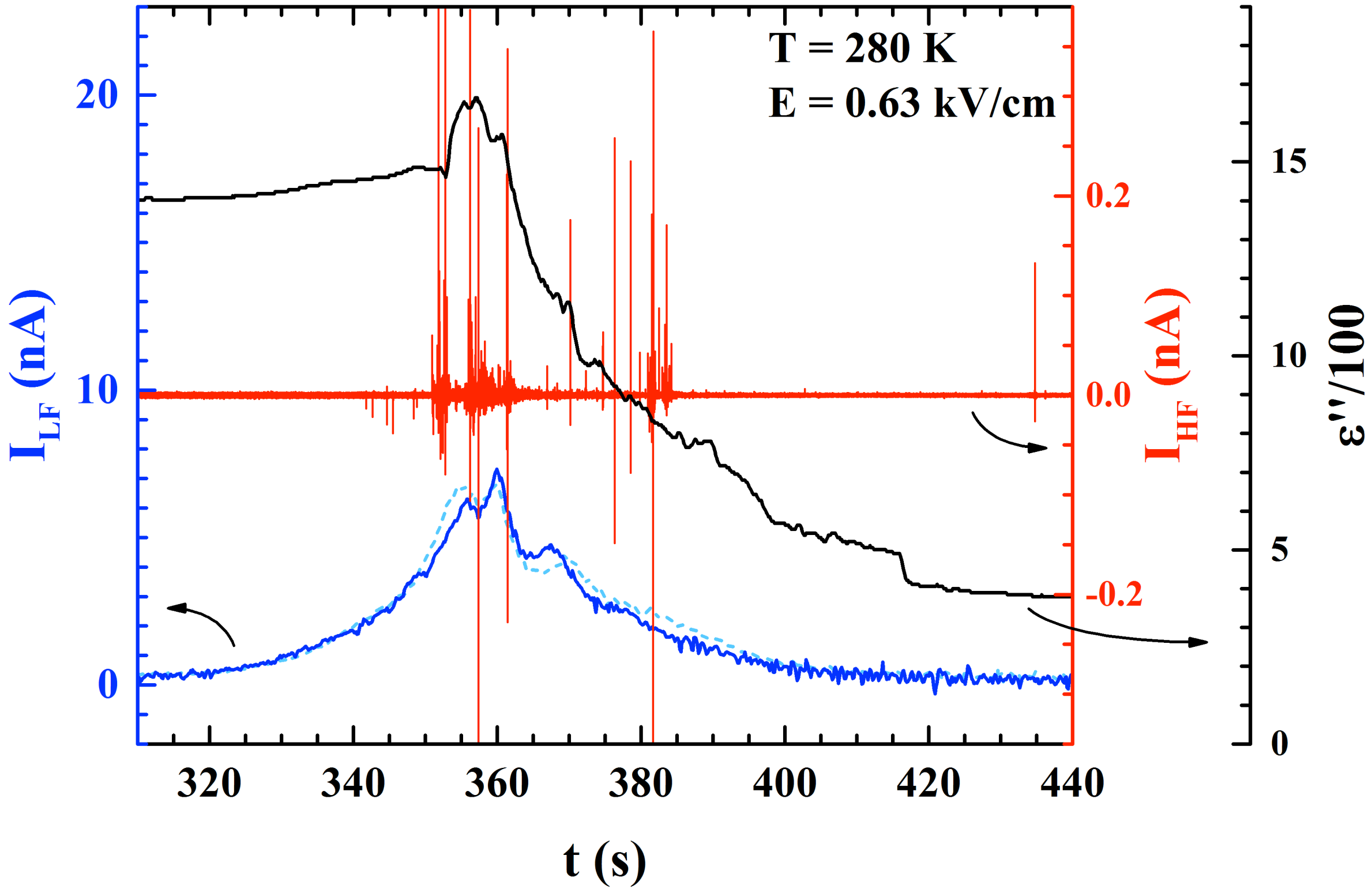

Fig. 8 shows HF noise and ε" (t), aligned via the LF $I_P(t)$ data. The solid $I_{LF}(t)$ goes with ε"(t) data and the dotted $I_{LF}(t)$ goes with $I_{HF}(t)$.